\documentstyle[11pt,aaspp4,flushrt]{article}

\tighten

\lefthead{Kriss et al.}
\righthead{ASCA X-ray Observations of NGC 3516}

\begin{document}

\title{ASCA Observations of the Composite Warm Absorber in NGC~3516}

\author{G. A. Kriss\altaffilmark{1},
J. H. Krolik\altaffilmark{1},
C. Otani\altaffilmark{2},
B. R. Espey\altaffilmark{1},
T. J. Turner\altaffilmark{3,4},
T. Kii\altaffilmark{5},
Z. Tsvetanov\altaffilmark{1},
T. Takahashi\altaffilmark{2},
A. F. Davidsen\altaffilmark{1},
M. Tashiro\altaffilmark{6},
W. Zheng\altaffilmark{1},
S. Murakami\altaffilmark{7},
R. Petre\altaffilmark{3},
and
T. Mihara\altaffilmark{2}
}

\altaffiltext{1}{Department of Physics \& Astronomy, Johns Hopkins University,
    Baltimore, MD 21218.}
\altaffiltext{2}{The Institute of Physical and Chemical Research (RIKEN),
    Hirosawa, Wako, Saitama 351-01, Japan.}
\altaffiltext{3}{Laboratory for High Energy Astrophysics,
    NASA/Goddard Space Flight Center, Greenbelt, MD 20771.}
\altaffiltext{4}{Universities Space Research Association.}
\altaffiltext{5}{Institute of Space and Astronautical Sciences,
    Yoshinodai, Sagamihara, Kanagawa 229, Japan.}
\altaffiltext{6}{Department of Physics, University of Tokyo, 7-3-1 Hongo,
    Bunkyo-ku, Tokyo 113, Japan.}
\altaffiltext{7}{Department of Physics, Osaka City University,
     3-3-138, Sugimoto, Sumiyoshi-ku, Osaka, 558, Japan.}

\authoremail{gak@pha.jhu.edu}

\begin{abstract}

We obtained X-ray spectra of the Seyfert 1 galaxy NGC~3516
in March 1995 using the Japanese X-ray satellite ASCA.
Simultaneous far-UV observations were obtained with
the Hopkins Ultraviolet Telescope on the Astro-2 shuttle mission.
The ASCA spectrum shows a lightly absorbed power law of energy index 0.78.
The low energy absorbing column is significantly less than previously seen.
Prominent {\sc O~vii} and {\sc O~viii} absorption edges are visible, but,
consistent with the much lower total absorbing column, no Fe K
absorption edge is detectable.
A weak, narrow Fe~K$\alpha$ emission line from cold material is present
as well as a broad Fe~K$\alpha$ line.  These features are similar to those
reported in other Seyfert 1 galaxies.
A single warm absorber model provides only an imperfect
description of the low energy absorption.
In addition to a highly ionized absorber with ionization parameter $U = 1.66$
and a total column density of $1.4 \times 10^{22}~\rm cm^{-2}$,
adding a lower ionization absorber with $U = 0.32$ and a
total column of $6.9 \times 10^{21}~\rm cm^{-2}$ significantly improves the
fit.
The contribution of resonant line scattering to our warm absorber models
limits the Doppler parameter to $< 160~\rm km~s^{-1}$
at 90\% confidence.
Turbulence at the sound speed of the photoionized gas
provides the best fit.
None of the warm absorber models fit to the X-ray spectrum can
match the observed equivalent widths of all the UV absorption lines.
Accounting for the X-ray and UV absorption simultaneously requires an
absorbing region with a broad range of ionization parameters
and column densities.

\end{abstract}


\section{Introduction}
Intrinsic absorption is a valuable tool
for probing structures in active galactic nuclei (AGN).
While absorption may in principle arise anywhere in the host galaxy, the
most interesting absorbers are those that appear to be associated with the
central engine.  These ``warm absorbers" commonly appear in the
X-ray spectra of AGN (\cite{Turner93}; \cite{NP94}), and they could be
material in the broad-emission-line region (BELR) (e.g. \cite{Netzer93};
\cite{RF95})
or the X-ray heated wind which forms the reflecting region in type 2 AGN
(\cite{KK95}), or an entirely new component.
If X-ray warm absorbers are related to associated UV absorption systems
(\cite{Mathur94}; \cite{Mathur95}), then UV and X-ray observations together
place powerful constraints on the ionization structure of the absorber.
In the X-ray one can measure the column densities of highly ionized species
(e.g. {\sc O~vii} and {\sc O~viii}) while simultaneously observing lower
ionization relatives in the UV ({\sc O~vi}, {\sc N~v}, and {\sc C~iv}).
Objects with strong UV absorption lines and soft X-ray absorption
are therefore good candidates for further tests of this hypothesis.

The Seyfert 1 galaxy NGC~3516 exhibits unusually strong, variable
UV absorption lines (\cite{UB83}; \cite{Voit87}; \cite{Walter90};
\cite{Kolman93}; \cite{Koratkar96}),
and has a variable X-ray spectrum characteristic of the warm absorber
phenomenon (\cite{Halpern82}).
Observations obtained with {\it Ginga} (\cite{Kolman93}; \cite{NP94})
show a flat power law with energy index $\sim 0.5$ over the 2--18 keV range,
a highly ionized iron edge with a corresponding total column density of
$\sim 3 \times 10^{23}~\rm cm^{-2}$,
and a cold fluorescent Fe K$\alpha$ line with EW = 377 eV.

To measure simultaneously the X-ray and UV absorption in NGC~3516
we coordinated ASCA observations with the flight of the Astro-2 space
shuttle mission in March 1995.  Far-ultraviolet spectra obtained with
the Hopkins Ultraviolet Telescope (HUT) that allow us to measure the
resonance doublets of {\sc O~vi}, {\sc N~v}, Si~{\sc iv} and {\sc C~iv}
are discussed in a companion paper by Kriss et al. (1996)\markcite{Kriss96a}.

\section{The ASCA Observations}

Four X-ray telescopes (\cite{Serlemitsos95})
consisting of nested conical foil mirrors image
X-rays onto four separate detectors in the ASCA focal plane.
Two solid-state imaging spectrometers (SIS0 and SIS1),
each consisting of four CCD chips, are sensitive from
$\sim 0.4$ to 10.0 keV and have an energy resolution of $\sim 2$\% at 6 keV
(\cite{Burke94}).
Two gas imaging spectrometers (GIS) cover the energy range 0.8--10.0 keV
with a resolution of $\sim 8$\% at 6 keV (\cite{Makishima96}).
The observatory and its performance are described by Tanaka, Inoue, \&
Holt (1994)\markcite{Tanaka94}.

We observed NGC~3516 twice, each time for roughly half a day with intervals
lost due to source occultation by the earth and passage through the
South Atlantic Anomaly (SAA).  The first observation was on 1995 March 11 from
02:24:39--14:40:48 UT.  The second was from 20:03:18 UT on 1995 March 12 to
08:11:00 UT on 1995 March 13.
The HUT spectra described by Kriss et al. (1996)\markcite{Kriss96a}
were obtained near the mid-point of each of these intervals.

\begin{figure}[b]
\plotfiddle{"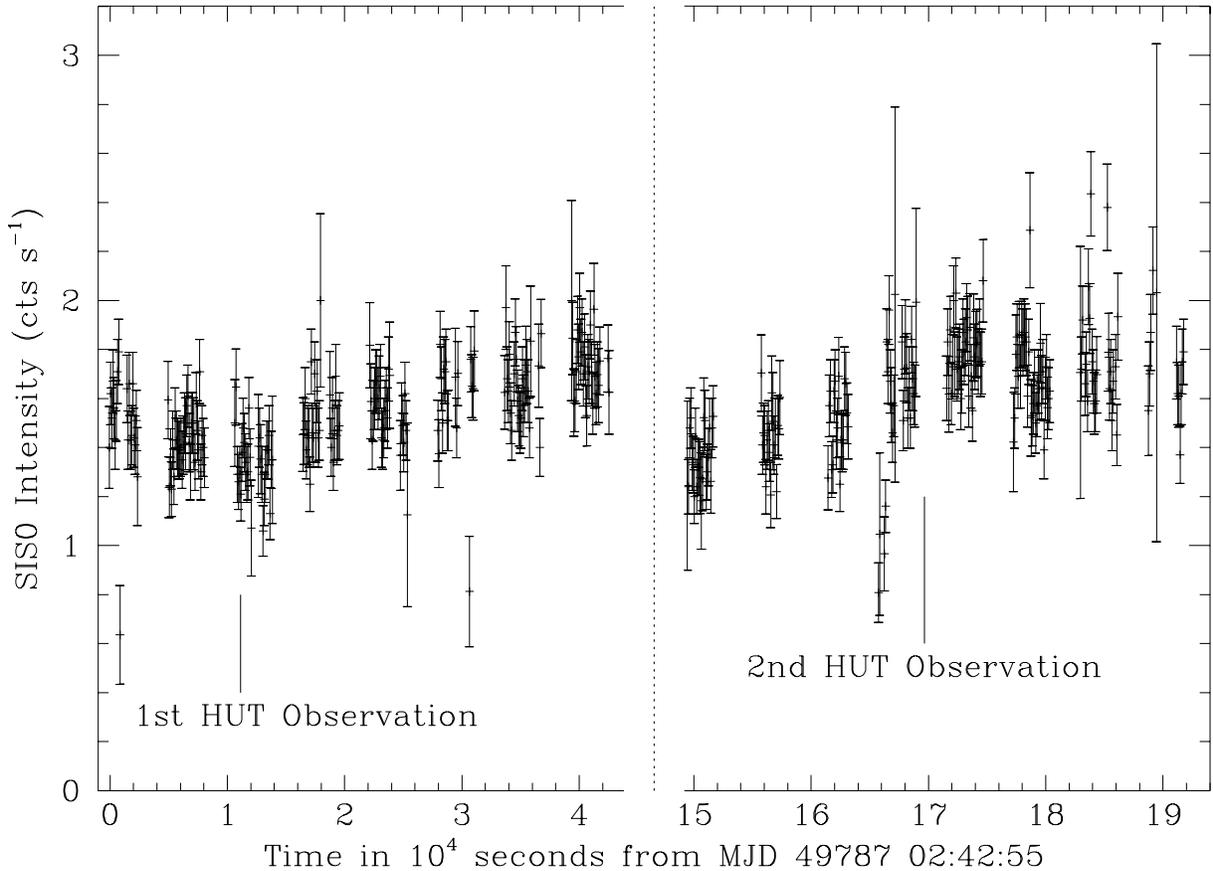"}{4.7in}{-90}{65}{65}{-255}{373}
\caption{
The integrated 0.4--10.0 keV count rate of NGC~3516 obtained with ASCA
is shown.  Time is in seconds from 02:42:55 UT on 11 March 1995.
Note the break in the X-axis between the first and second observations.
Data are binned into intervals of 100~s.  Variations of $\sim 20$\% are
apparent on timescales of $\sim 20,000~\rm s$.
The centroids of the integration intervals for the two HUT spectra
reported by Kriss et al. (1996) are indicated.
\label{n3516x1pp.ps}}
\end{figure}

The SIS detectors were operated in FAINT 1-CCD mode, but
the spectra extracted in FAINT mode showed regions of large pixel-to-pixel
variations.  Converting the data to BRIGHT mode eliminated these
variations, and we used these spectra for the analysis presented here.
Experimentation with different data screening criteria showed that
bright earth avoidance angles of $> 25$\arcdeg\ were necessary to achieve
consistent results.
Default screening criteria were used for the dark earth limb angle
($> 10$\arcdeg) and for the geomagnetic cutoff rigidity ($> 6~\rm GeV~c^{-1}$).
A few high count rate intervals adjoining SAA
passages were rejected by eliminating regions with count rates exceeding
$2.5~\rm cts~s^{-1}$.
Counts for light curves and spectra were extracted from circular
regions centered on NGC~3516 using the largest radius that
did not extend beyond the boundary of the CCD (3\farcm5 for SIS0 and 2\farcm6
for SIS1).
As shown by the SIS0 count rate in Figure \ref{n3516x1pp.ps},
NGC~3516 showed intensity changes of
$\sim 20$\% over the course of both observations.
The combined data sets yield 54,539 events over a live integration time
of 33,534 s in SIS0 and 38,935 events over 33,956 s in SIS1.

\section{Modeling the X-ray Spectrum}

We used the spectral fitting program XSPEC (\cite{xspec89})
to model the extracted spectra.  To permit the use of $\chi^2$ statistics
we grouped the data to have a minimum of 25 counts per spectral bin.
This affected only bins above 6.5 keV.
To avoid the worst uncertainties in the detector response, we restricted
our fits to bins with energies $0.60~\rm keV~< E < 9.00~keV$.
The SIS0 and SIS1 data were fit jointly.

Although spectral features are obvious in the raw spectrum, we first fit
a simple power law with cold absorption to the data to draw a comparison
with earlier X-ray observations.  This model yields an energy index of 0.62,
a flux at 1 keV of $0.01~\rm photons~cm^{-2}~s^{-1}~keV^{-1}$,
an equivalent neutral hydrogen column of
$N_H = 1.2 \times 10^{21}~\rm cm^{-2}$ and $\chi^2 / \nu = 1158/ 435$.
The spectral shape and intensity is significantly different from that found
in earlier {\it Ginga} observations (\cite{Kolman93}).
The intensity as observed with ASCA is about twice as high,
the spectral index is steeper ($0.62 \pm 0.03$ compared to a mean of
$0.47 \pm 0.04$ in the {\it Ginga} data), and the absorbing column
$\sim 30 \times$ lower than the value of
$\sim 4.06 \pm 0.28 \times 10^{22}~\rm cm^{-2}$
found by Kolman et al. (1993)\markcite{Kolman93}.
As {\it Ginga} is sensitive to column densities of
$\sim 1 \times 10^{21}~\rm cm^{-2}$, there is little doubt that the
low energy absorption has changed significantly in character since 1989.
Such striking differences in column density were seen previously in the
{\it Einstein} observations of NGC~3516 (\cite{Halpern82}), and
they are typical of variations seen in other sources with intrinsic
absorption such as NGC~4151 (\cite{Yaqoob89}; \cite{Yaqoob93})
and MR2251-178 (\cite{Halpern84}; \cite{Otani96b}).
The spectral index, however, is sensitive to the modeling of
the absorption and the iron emission in the spectrum.
Using the same {\it Ginga} data, Nandra \& Pounds (1994)\markcite{NP94}
find a spectral index of 0.65--0.74 and absorption columns of
$5.0-7.2 \times 10^{22}~\rm cm^{-2}$.

\begin{figure}[t]
\plotfiddle{"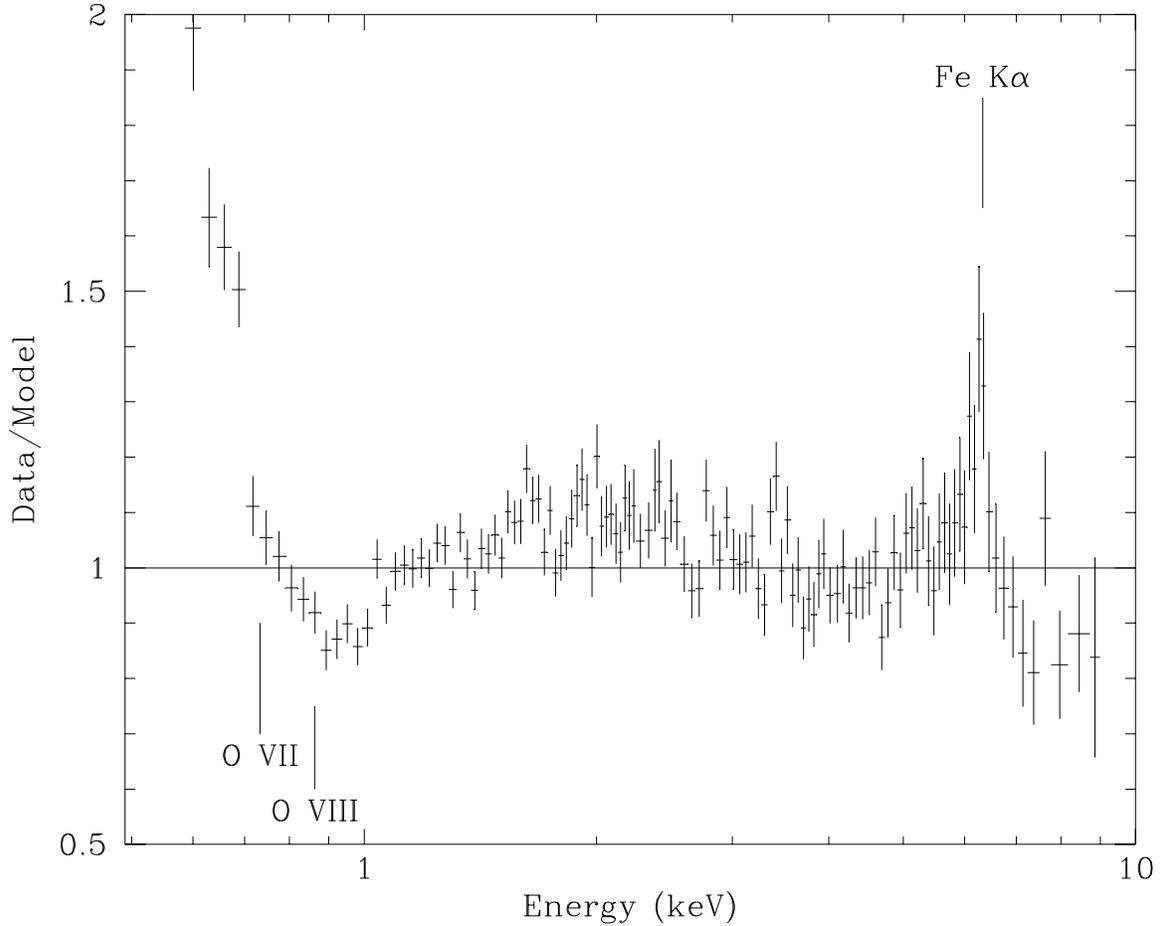"}{4.7in}{-90}{65}{65}{-255}{373}
\caption{
To illustrate the features present in the X-ray spectrum of
NGC~3516, the data were divided by a simple model consisting of a power law
with
low energy absorption by neutral gas.  The ratio of the data
to the model shows a prominent absorption dip around the photoionization
edges of {\sc O~vii} and {\sc O~viii} as well as complex structure
around the iron K$\alpha$ line.
\label{n3516x2pp.ps}}
\end{figure}

The ratio of the SIS0 data to the simple power-law model shown
in Figure \ref{n3516x2pp.ps}
immediately reveals absorption features below 1 keV due to
highly ionized oxygen and strong fluorescent emission from neutral iron
around 6.3 keV.
To characterize these spectral features empirically, we added a succession
of Gaussian-profile emission lines and absorption edges (with opacity
proportional to $E^{-3}$) until we obtained an acceptable fit.
An acceptable description of the spectrum requires two low
energy absorption edges that we attribute to {\sc O~vii} and {\sc O~viii},
a narrow unresolved Fe K$\alpha$ emission line,
and a broad base to the Fe K$\alpha$ line.
Table \ref{empirical_table} gives the best fit values and 90\% confidence
error bars for the parameters of this empirical model.

The broad and narrow Fe K$\alpha$ features have a combined equivalent width
(EW)
of $253 \pm 109$ eV, comparable to the $377 \pm 44$ eV Fe K$\alpha$
feature in the {\it Ginga}
spectrum of Kolman et al. (1993)\markcite{Kolman93}.
The energies of the oxygen edges and the narrow Fe K$\alpha$ line are
redshifted
relative to the systemic redshift of NGC~3516 ($z = 0.008834$, \cite{VC85}).
The {\sc O~vii}, {\sc O~viii}, and Fe K$\alpha$ features have redshifts of
$z = 0.046 \pm 0.011$, $z = 0.019 \pm 0.018$, and $z = 0.016 \pm 0.0064$,
respectively.
Inclusion of the two oxygen edges in the fit broadens the distribution of
opacity at low energies.
This more accurate modeling of the shape of the low energy absorption
leads to a spectral index of 0.78 that is steeper than that in the simple
power-law fit and is more comparable to the mean power law index of 0.73
found for {\it Ginga} observations of Seyferts (\cite{NP94}).
The residual absorption by cold gas of
$N_{HI} = 6.8 \times 10^{20}~\rm cm^{-2}$, however, is still higher than
the expected Galactic column of $N_H = 3.35 \times 10^{20}~\rm cm^{-2}$
(\cite{Stark92}).
At first glance it might seem natural to attribute this
excess above the Galactic column to cold gas intrinsic to NGC~3516.
However, the far-UV spectrum obtained with HUT
limits the intrinsic neutral hydrogen column in NGC~3516 to
$< 8 \times 10^{17}~\rm cm^{-2}$ (\cite{Kriss96a})
based on the strengths of the observed Lyman lines and Lyman limit.
The most likely explanation for the observed excess cold column is the
uncertainty in the ASCA calibration below 1 keV.
These uncertainties can lead to excess column
densities of up to $3 \times 10^{20}~\rm cm^{-2}$.

\begin{deluxetable}{ l c }
\tablecolumns{2}
\tablewidth{468pt}
\tablecaption{Empirical Model Parameter Values\label{empirical_table}}
\tablehead{
\colhead{Parameter} & \colhead{Value\tablenotemark{a}}\\
}
\startdata
$N_{HI}$~$(10^{20}~\rm cm^{-2})$\tablenotemark{b} & $6.8 \pm 1.0$ \nl
Edge Energy $E_1$ (keV) & $0.707 \pm 0.008$ \nl
Optical Depth $\tau_1$ & $0.65 \pm 0.07$ \nl
Edge Energy $E_2$ (keV) & $0.855 \pm 0.015$ \nl
Optical Depth $\tau_2$ & $0.31 \pm 0.05$ \nl
Energy index $\alpha$  & $0.776 \pm 0.029$ \nl
Normalization ($\rm phot~cm^{-2}~s^{-1}~keV^{-1}$) & $0.0126 \pm 0.00037$ \nl
Narrow Fe Energy (keV)\tablenotemark{c} & $6.293 \pm 0.040$ \nl
Narrow Fe EW (eV)      & $73 \pm 27$ \nl
Broad Fe Energy (keV) & $5.88 \pm 0.37$ \nl
Broad Fe EW  (eV)     & $180 \pm 106$ \nl
Broad Fe FWHM (keV)   & $1.58 \pm 0.84$ \nl
$\chi^2 / \nu$ & 494.7/426\tablenotemark{d} \nl
\enddata
\tablenotetext{a}{
The quoted errors are 90\% confidence for a single interesting parameter.}
\tablenotetext{b}{
Equivalent neutral hydrogen column for absorption by cold (neutral) gas.}
\tablenotetext{c}{
The intrinsic width of this unresolved line was held fixed at 0 eV.
The 90\% confidence upper limit on the line FWHM is 39 eV.}
\tablenotetext{d}{
The probability of exceeding $\chi^2$ in this fit is 0.012.}
\end{deluxetable}

Iron edges are also important diagnostics of the ionization state of the
absorbing medium.  No iron edge feature is apparent in the residuals from
our fit, and this is not surprising given the order-of-magnitude
lower columns we are seeing relative to earlier X-ray observations of
NGC~3516.  Adding an additional sharp edge with its energy constrained to be
greater than 7.1 keV gives no improvement in $\chi^2$.
The optical depth at the edge must be less than 0.3 at the 90\%
confidence level.

The moderately strong edges of highly ionized oxygen in our spectrum
naturally suggest an origin in photoionized gas.
The dominant strength of the {\sc O~vii} edge indicates gas of much lower
ionization and temperature than that modeled by
Krolik \& Kriss (1995)\markcite{KK95},
and the absence of an Fe K edge indicates a much smaller column.
Accordingly we have computed new models that span the potential range
of ionization parameters and column densities with some slight modifications
to the procedure described by Krolik \& Kriss (1995)\markcite{KK95}.
First, we used an ionizing spectrum appropriate for NGC~3516 at this epoch.
The best-fit UV power law of $f_\nu \sim \nu^{-1.89}$ (\cite{Kriss96a}) was
extrapolated to higher frequencies, and the best-fit X-ray power law found here
$f_\nu \sim \nu^{-0.78}$ was extrapolated to lower frequencies;
the two meet at 51 eV.
Second, the lower temperatures and ionization state place the gas in a
regime where thermal equilibrium is more likely because the cooling time
is rather shorter.
We therefore compute our models in thermal equilibrium.
Finally, for ease of comparison to warm absorber models fit to the X-ray
spectra of other AGN, we assume constant density clouds and use the ionization
parameter $U = n_{ion} / n_H$, where $n_{ion}$ is the number density of
ionizing photons between 13.6 eV and 13.6 keV illuminating the cloud and
$n_H$ is the density of hydrogen atoms.\footnote{
For comparison to the constant-pressure models of Krolik \& Kriss (1995), the
ionization parameter $\Xi = P_{rad} / P_{gas} = 5.3~U$ for gas at $10^5$ K.}
The transmission of each model is computed exactly as described by
Krolik \& Kriss (1995)\markcite{KK95}, taking into account resonant line
scattering and electron scattering as well as continuum opacity.
The resulting models are a two parameter family in total column density $N$ and
ionization parameter.  These are read into XSPEC as a FITS table for
fitting to the X-ray spectrum.

As in Krolik \& Kriss (1995)\markcite{KK95} we assume low density gas
($n_H = 10^3~\rm cm^{-3}$), but there are no density-dependent effects in our
calculations or results, provided one considers densities lower than
$\sim 10^{11}~\rm cm^{-3}$.
Although the UV continuum in our photoionizing spectrum is steep, we note
that there is no lack of high energy photons.  The spectrum flattens
just below the He~{\sc ii} edge to $f_\nu \sim \nu^{-0.78}$,
and the overall $\alpha_{ox} = 1.20$,
a value typical of Seyfert 1 galaxies (Kriss \& Canizares 1985)\markcite{KC85}.
In fact, tests show that our results are rather insensitive to the exact
shape of the ionizing spectrum apart from changes in the deduced ionization
parameter.
(This is a general property of photoionization models with broad distributions
of ionizing flux noticed in even the
earliest calculations [Tarter, Tucker, \& Salpeter 1969]\markcite{TTS69}.)
For comparison we computed alternative models assuming either an extremely hard
spectrum with $f_\nu \sim \nu^{-0.78}$ from 2500 \AA\ through
the UV and X-ray up to 100 keV,
or the spectral shape of NGC~5548 as used by
Mathur et al. (1995)\markcite{Mathur95}, which contains a soft X-ray excess.
Neither of these match the observed broad-band spectral shape, yet they
both provide equally good descriptions of the opacity of the warm absorber.

The Fe K$\alpha$ lines in our empirical fit to the ASCA spectrum
are indicative of fluorescent emission.
As a better model for the X-ray continuum shape we therefore
use that predicted by Lightman \& White (1988)\markcite{LW88}
for a disk illuminated by a power law.
Our data do not constrain the inclination or the solid angle
subtended by the disk, so we fix these parameters at 30\arcdeg\ and $2 \pi$,
respectively.  The source is assumed to radiate isotropically, and we impose
a high energy cutoff of 300 keV on the intrinsic power law.

Using the optical depths given by our best empirical fit in
Table \ref{empirical_table} and
the threshold photoionization cross sections of Verner \& Yakovlev
(1995)\markcite{VY95}, we infer column densities for {\sc O~vii} and
{\sc O~viii} of $3.5 \times 10^{18}~\rm cm^{-2}$ and
$2.6 \times 10^{18}~\rm cm^{-2}$, respectively.  Assuming these represent
all the oxygen atoms, the equivalent total hydrogen column for a solar
abundance of oxygen is $9.3 \times 10^{21}~\rm cm^{-2}$.
Replacing the two oxygen edges in our empirical model with the grid of warm
absorber models, for the best fit we obtain a column density and ionization
parameter that qualitatively matches our expectations,
as shown in the center column of Table \ref{warmabs_table}.
This fit is only slightly worse than our empirical model.
Warm absorber models computed with our alternative ionizing spectra give
identical best fit values for the total column density
($\log N = 22.02$).  For the hard $\nu^{-0.78}$ spectrum the best-fit
ionization
parameter is $U = 0.25$ with $\chi^2 / \nu = 501.5 / 427$, and
for the spectrum like NGC~5548, $U = 1.71$ with $\chi^2 / \nu = 500.7 / 427$.

\begin{deluxetable}{ l c c }
\tablecolumns{3}
\tablewidth{468pt}
\tablecaption{Warm Absorber Model Parameters\label{warmabs_table}}
\tablehead{
\colhead{Parameter} & \colhead{Single Absorber\tablenotemark{a}} & \colhead{Two
Absorbers\tablenotemark{a}}\\
}
\startdata
$N_{HI}$~($10^{20}~\rm cm^{-2}$)\tablenotemark{b} & $7.1 \pm 1.7$ & $7.7 \pm
1.6$ \nl
Ionization Parameter $U_1$     & $0.48 \pm 0.06$ & $1.66 \pm 0.31$ \nl
log Total Column Density $N_1$ & $22.02 \pm \phantom{0}0.06$  & $22.15 \pm
\phantom{0}0.12$ \nl
Redshift $z_1$                 & $0.037 \pm 0.011$ & $0.007 \pm 0.014$ \nl
Ionization Parameter $U_2$     & ...               & $0.32 \pm 0.11$ \nl
log Total Column Density $N_2$ & ...               & $21.84 \pm
\phantom{0}0.04$ \nl
Redshift $z_2$                 & ...               & $0.050 \pm 0.013$ \nl
Intrinsic Energy index $~\alpha$\tablenotemark{c}  & $0.97 \pm 0.08$ & $1.05
\pm 0.07$ \nl
Normalization ($\rm phot~cm^{-2}~s^{-1}~keV^{-1}$) & $0.0167 \pm 0.0014$ &
$0.0191 \pm 0.0020$ \nl
Narrow Fe Energy (keV)\tablenotemark{d} & $6.289 \pm 0.039$ & $6.294 \pm 0.038$
\nl
Narrow Fe EW (eV)      & $76 \pm 29$ & $79 \pm 30$ \nl
Broad Fe Energy (keV) & $5.91 \pm 0.33$ & $6.00 \pm 0.34$ \nl
Broad Fe EW (eV)      & $387 \pm 170$ & $427 \pm 171$ \nl
Broad Fe FWHM (keV)   & $2.33 \pm 0.63$ & $2.47 \pm 0.89$ \nl
$\chi^2 / \nu$ & 500.7/427\tablenotemark{e} & 486.5/424\tablenotemark{f} \nl
\enddata
\tablenotetext{a}{
The quoted errors are 90\% confidence for a single interesting parameter.}
\tablenotetext{b}{
Equivalent neutral hydrogen column for absorption by cold (neutral) gas.}
\tablenotetext{c}{
This is the intrinsic energy index for a power-law spectrum illuminating a
cold disk of gas.  The inclination of the disk is fixed at 30\arcdeg\ and its
covering fraction is fixed at $2 \pi$.}
\tablenotetext{d}{
The intrinsic width of this unresolved line was held fixed at 0 eV.}
\tablenotetext{e}{
The probability of exceeding $\chi^2$ in this fit is 0.0079.}
\tablenotetext{f}{
The probability of exceeding $\chi^2$ in this fit is 0.019.}
\end{deluxetable}

\begin{figure}[t]
\plotfiddle{"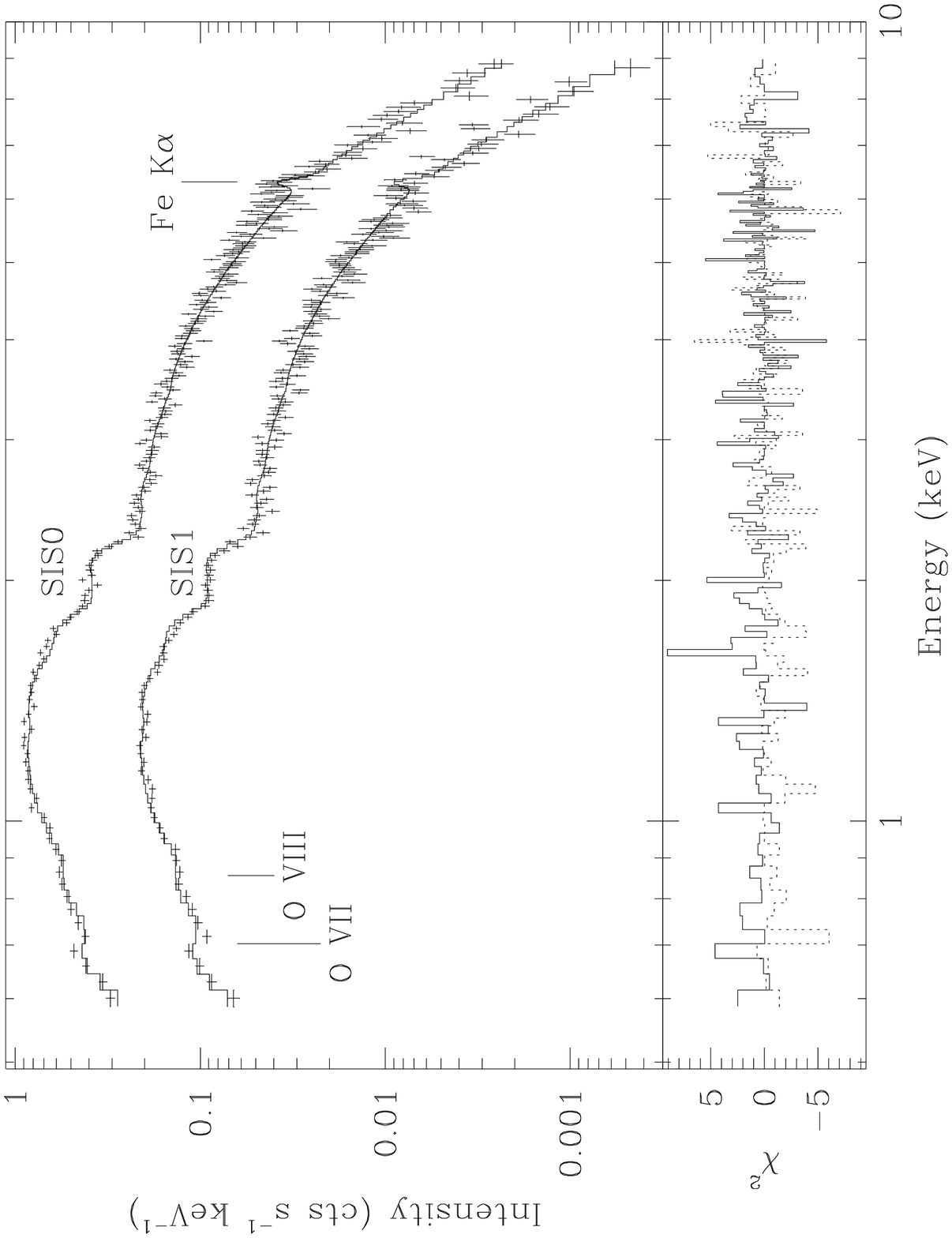"}{4.7in}{-90}{65}{65}{-255}{373}
\caption{
{\it Upper Panel:} The solid lines are the best-fit empirical model folded
through the ASCA SIS0 and SIS1 detector responses.
The data points are crosses with 1$\sigma$ error bars.
The SIS1 data are offset down by 0.5 in the log for clarity.
The model includes a power law with photon index 1.78, absorption by
neutral gas with an equivalent neutral hydrogen column of
$\rm N_H = 6.8 \times 10^{20}~cm^{-2}$,
a photoionization edge due to {\sc O~vii} at 0.71 keV with an optical
depth at the edge of 0.65,
a photoionization edge due to {\sc O~viii} at 0.86 keV with optical depth 0.31,
an unresolved iron K$\alpha$ line at 6.29 keV with an equivalent width of 73
eV,
and a broad (FWHM = 1.58 keV) iron K$\alpha$ line at 5.88 keV with an
equivalent width of 180 eV.
{\it Lower Panel:}
The contributions to $\chi^2$ of each spectral bin are shown.
The solid line is for SIS0 and the dotted line for SIS1.
\label{n3516x3pp.ps}}
\end{figure}

As discussed in our companion paper (\cite{Kriss96a}), the strengths of the
UV lines observed in NGC~3516 require a zone of lower ionization and lower
column density than this warm-absorber model.
Other observations also indicate that the warm absorbing medium may be
complex.
Otani et al. (1996)\markcite{Otani96} find that the {\sc O~viii} opacity
in MCG--6-30-15 is variable, while the {\sc O~vii} opacity is not,
suggesting that the absorption arises in at least two different zones.
Prompted by these suggestions of spectral complexity, we experimented with
adding a second warm absorber to the fit.
This significantly improves $\chi^2$.
Although the level of improvement is not sufficient to be apparent in any
particular feature in the spectrum,
an F test for 3 additional parameters producing $\Delta \chi^2 = 14.2$
shows that this second model component is significant at the 99\%
confidence level.
The parameters of this best fit are summarized in the last column
of Table \ref{warmabs_table}.
The SIS0 and SIS1 spectra and this best fit model are illustrated in
the top panel of Figure \ref{n3516x3pp.ps}.

\begin{figure}[t]
\plotfiddle{"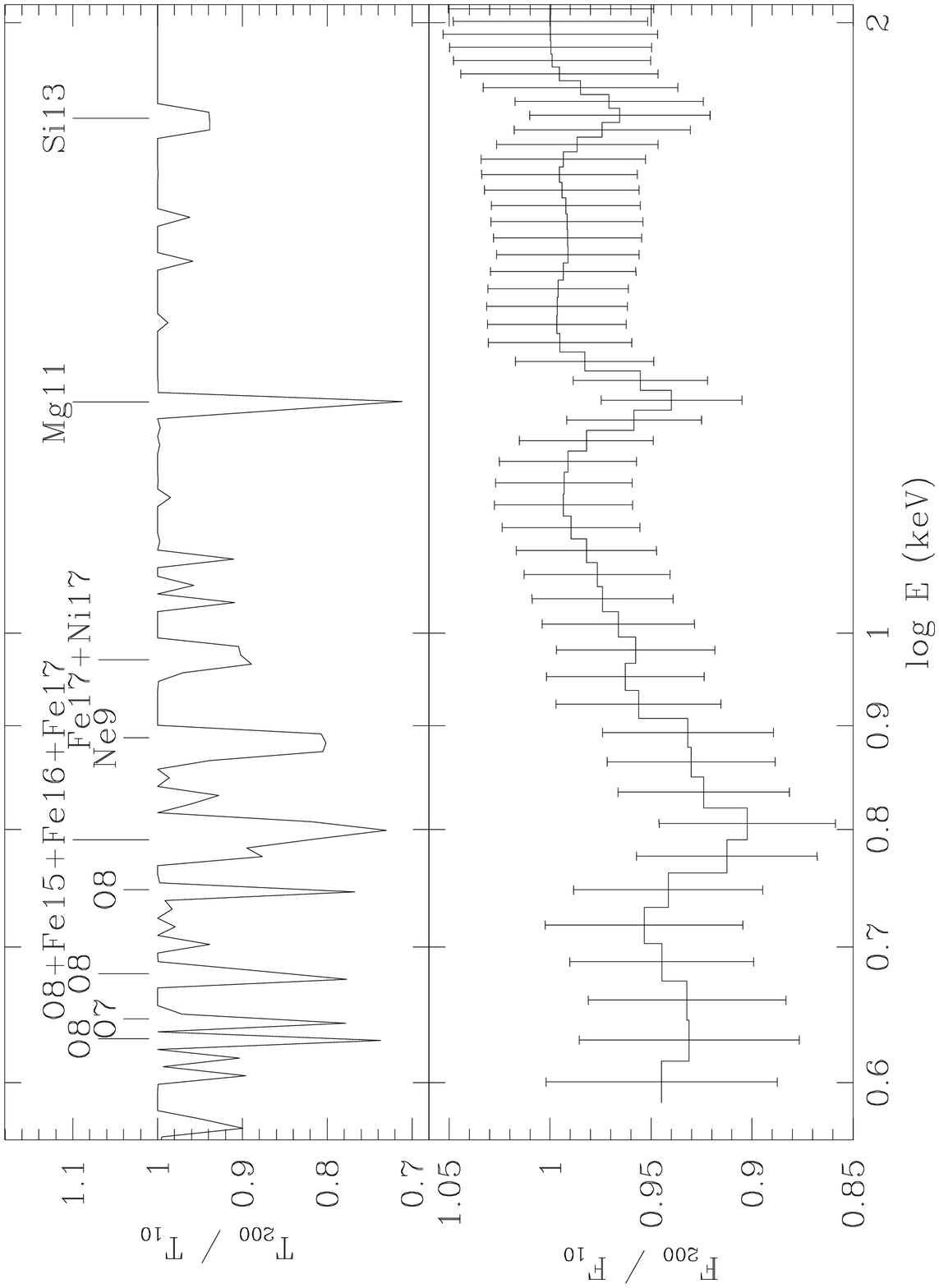"}{4.7in}{-90}{65}{65}{-255}{373}
\caption{
{\it Top Panel:}
The ratio of the transmission for a warm absorber model with absorption
lines broadened by a Doppler parameter of $b = 200 \rm~km~s^{-1}$ to
the same model broadened with $b = 10 \rm~km~s^{-1}$.
Physical parameters as determined by the best-fit single absorber model
in Table 2 were used for the ionization parameter and
column density.
Ion species contributing to the most significant lines are marked.
{\it Lower Panel:}
Here the models are folded through the ASCA response function before
computing the ratio.
The error bars on each bin are taken from the corresponding data at that point.
\label{n3516x4pp.ps}}
\end{figure}

As in Krolik \& Kriss (1995)\markcite{KK95}
all the warm absorber models above assumed resonant line scattering profiles
with Doppler parameters given by the sound speed in the photoionized gas.
Since the line opacity is sensitive to the assumed profile width, it
influences the transmission computed for each model.
At low velocities the opacity is dominated by continuum absorption;
at high velocities resonant line scattering makes a significant contribution.
To illustrate, Figure \ref{n3516x4pp.ps} shows the computed
transmission for $b = 200 \rm~km~s^{-1}$ divided by that for
$b = 10 \rm~km~s^{-1}$.
Physical parameters as determined by the best-fit single absorber model
in Table \ref{warmabs_table} were used for the ionization parameter and
column density.
The upper panel shows the ratio of the models themselves.
The lower panel shows the ratio after the models have been folded through
the ASCA response function.  The error bars on each bin are taken from the
corresponding data points.

The stronger predicted line absorption at $b = 200 \rm~km~s^{-1}$
is clear.  Below 1 keV {\sc O~viii} and Fe-L transitions dominate the
increased opacity.  The only significant features above 1 keV are the resonance
transitions of Mg~{\sc xi} and Si~{\sc xiii}.
At $b = 10 \rm~km~s^{-1}$ the absorption lines have an integrated
equivalent width of 5 eV and make a negligible contribution to the total
opacity, whereas at $b = 200 \rm~km~s^{-1}$ their equivalent width is 42 eV
and 6\% of the opacity between 0.6 and 2.0 keV is due to line absorption.

Within the context of our photoionization models the dependence of line opacity
on line width permits us to constrain the Doppler parameter, even though
discrete absorption lines are not
unambiguously present in the observed spectrum.
In essence we find that the {\it absence} of significant resonant absorption
line features permits us to set upper limits on the Doppler parameter.
We computed grids of models with Doppler parameters varying from
$10~\rm km~s^{-1}$ to $200~\rm km~s^{-1}$.
The best fits for both the single and the double warm absorber have
Doppler parameters of $50~\rm km~s^{-1}$, approximately the sound
speed for these models.
(For the double warm absorber model we used the same Doppler parameter
for each component.)
At 90\% confidence for a single interesting
parameter ($\Delta \chi^2 = 2.706$) we constrain the Doppler parameter
to values less than 160 $\rm km~s^{-1}$ for the single warm absorber model,
and to less than 120 $\rm km~s^{-1}$ for the model with two warm absorbers.

\section{Discussion}

The broad Fe K$\alpha$ emission line in our spectrum of NGC~3516
resembles features seen in MCG--6-30-15 (\cite{Tanaka95}; \cite{Fabian95}),
in NGC~5548 and IC~4329A (\cite{Mushotzky95}) and
in NGC~4151 (\cite{Yaqoob95}).
At the signal-to-noise ratio of our spectrum we are unable to place
significant constraints on relativistic disk models.
In fact, the width and luminosity of the line depends strongly on
our model for the underlying continuum.
In the empirical model, summarized in Table \ref{empirical_table}, the line
width is smaller largely because the underlying continuum is flatter and
has less curvature than in the warm absorber models.
In the warm absorber models the widths and equivalent widths are large,
but given the uncertainties, they are compatible with the maximum equivalent
widths of $\sim 200$ eV expected in models of X-ray reflection from
cold disks (e.g., \cite{GF91}).

A simple, single-zone photoionized absorber is an imperfect description
of the soft X-ray opacity in NGC~3516.
At least two zones are required to give as
good a fit to the data as our empirical model containing discrete absorption
edges due to {\sc O~vii} and {\sc O~viii}.
These two zones may be a simplification of a broad distribution of
ionization parameters in the absorbing gas, or they may be indicative
of two entirely different regions as suggested by
Otani et al. (1996)\markcite{Otani96} in their study of the variability
of {\sc O~vii} and {\sc O~viii} opacity in MCG--6-30-15.
The complexity of the absorbing gas in NGC~3516 increases even more when one
considers the UV absorption lines observed simultaneously with HUT.
In the companion to this paper Kriss et al. (1996)\markcite{Kriss96a} find
that neither a single photoionized absorber nor the multiple warm absorber
models considered here are adequate to explain the
UV absorption lines.
This confirms the conclusion of Kolman et al. (1993)\markcite{Kolman93}
that there is probably not a direct connection between the warm X-ray
and the UV absorbers in NGC~3516.

Having separate regions for the X-ray and UV absorption
contrasts with the conclusions of
Mathur et al. (1994,1995)\markcite{Mathur94}\markcite{Mathur95}
in their studies of the warm absorbers in 3C~351 and in NGC~5548.
For those objects they described a single absorbing zone that could
account for both the X-ray warm absorber and the associated UV absorption
lines.
Not all X-ray and UV absorbers may be so simple.
A clue to the additional complexity of the absorption in NGC~3516 is the
presence of large columns of lower ionization species such as
Si~{\sc iv} and an optically thick Lyman limit.
Low ionization species also present problems for single zone models
in attempting to model UV and X-ray absorption in NGC~4151 (\cite{Kriss95}).
Since Si~{\sc iv} absorption is strong in broad absorption line quasars,
by analogy this may indicate that single zone models will also not
suffice to explain both the UV and X-ray absorption in these objects,
contrary to the suggestion of Mathur, Elvis, \& Singh (1995).\markcite{MES}

Although the absorbing medium in NGC~3516 appears to be highly stratified,
the presently observed weak X-ray absorption coincident with an episode of
weak UV absorption suggests that some underlying mechanism ties them together.
At the epoch of our observation, NGC~3516 was brighter than usual by about a
factor of two,
but the low energy X-ray absorption
decreased by more than can be accounted for simply
by photoionization due to the increased luminosity.
As suggested by Walter et al. (1990)\markcite{Walter90},
these large changes in absorption column may be caused
by different clouds moving across the line of sight.
Rather than a single cloud, however, we require a
whole population of clouds of differing column densities and ionization
parameters moving into place.
If the absorption arises in a wind driven from the accretion disk or the
obscuring torus, large changes in opacity that are correlated in the UV and the
X-ray may be linked by fluctuations in the mass supply to the outflow.
X-ray and UV absorption in AGN may ultimately have a common origin, but
the absorption probably occurs in distinctly different regions with a variety
of
physical conditions.

\acknowledgments

This work was supported by a NASA ASCA guest investigator grant NAG 5-2935 and
a NASA Long-Term Space Astrophysics grant NAGW-4443 to the Johns Hopkins
University.
CO thanks the Special Postdoctoral Researchers Program of RIKEN for support.



\end{document}